\documentclass{article}[11pt]
\usepackage{graphicx} 
\usepackage{amssymb,enumerate,graphicx,verbatim,amsthm,bbm}
\usepackage{todonotes}
\usepackage[inline, shortlabels]{enumitem}
\usepackage{thmtools}
\usepackage{thm-restate}
\usepackage[hidelinks]{hyperref}
\usepackage{subcaption}
\usepackage{mwe}
\usepackage[toc,page,header]{appendix}
\usepackage{minitoc}
\usepackage{algorithmic}
\usepackage[linesnumbered,ruled,vlined]{algorithm2e}
\usepackage[left=3.0cm, right=3.0cm, top=3cm]{geometry}
\usepackage{amsmath}
\usepackage[nameinlink,noabbrev,capitalise]{cleveref}
\Crefname{table}{Table}{Tables}
\Crefname{assumption}{Assumption}{Assumptions}
\crefname{equation}{}{}

\usepackage{varwidth}
\usepackage{authblk}
\usepackage[numbers]{natbib}

\title{Bayesian Predictive Probabilities for Online Experimentation}

\author{Abbas Zaidi}
\author{Rina Friedberg}
\author{Samir Khan}
\author{Yao-Yang Leow}
\author{Maulik Soneji}
\author{Houssam Nassif}
\author{Richard Mudd}
\affil{Meta Inc \authorcr
  \{\tt \normalsize abbaszaidi, rinafriedberg, samirk, leowyaoyang, soneji, houssamn, rmudd\}@meta.com}
\date{}
\date{}

\begin{document}

\maketitle
\section{Introduction}
Online Randomized Controlled Experiments (A/B Tests) enable product decision-making across the information technology sector~\citep{Fiez2024Anduril}. Among a myriad of applications, they are used to support deployment decisions and to evaluate the impact of product changes and feature launches  \citep{barajas2021online}.
The design may vary significantly across experiments, and various organizations (even at the same company) may follow different processes - e.g., some may perform replication experiments while others do not. Despite these variations, product decision processes ultimately depend on a company's ability to perform high-quality inference across its collection of A/B tests \citep{kohavi2020trustworthy}.

However, given the role of experimentation in the product life-cycle, finite resources present an evergreen challenge in this space; compute (e.g., online machine usage) and experimental units (e.g., finite number of users to advertise to) being chief amongst such constraints. Downstream of said limitation, users of experimentation platforms are increasingly motivated to use ad-hoc means of optimizing these resources primarily by way of \textit{peeking} \citep{johari2017peeking}. Their objective is to conclude futile experiments or make decisions from clearly efficacious ones cheaply and quickly given the omnipresence of limited resources. As an example, ongoing monitoring of trends in point-estimates and uncertainty intervals is a common operationalization of this practice. However, the absence of formalized stopping rules~\citep{stallard2001stopping} are error prone and often strongly misaligned with end-of-experiment outcomes, leading to inflated type-I errors and erroneous conclusions~\citep{maharaj2023anytime, Weltz2023heteroskedastic}.

This work introduces a mechanism for principled interim-analysis using \textit{Bayesian Predictive Probabilities}~\citep{saville2014utility} that have been widely utilized in applications outside of the technology domain. These methods enable statistically efficient, data-driven decision making from experiments that are robust to peeking by way of having a stopping rule that is \textit{ignorable}~\citep{gelman1995bayesian}.  

Our work contributes to the growing body of research on interim analyses in online experimentation in two ways. First, our approach introduces a novel application of predictive probabilities in online experimentation, which to our knowledge have not been utilized in this domain before. Second, we demonstrate how this system can be deployed at scale as a means of improving the efficiency of online experimentation systems without complex inference schemes. As a result, we expect the proposed approach to be a viable mechanism for practitioners to more optimally utilize these methods for large-scale online experimentation platforms.

\section{Methods}
\label{sec:methods}
\subsection{Notation}
Assume that at the conclusion of the experiment, the treatment under consideration is declared a success if $\mathrm{P}(\theta > 0|\hat{\theta}_{1,\ldots, T}) > (1-\alpha)$. Here $\theta$ denotes the true but unknown effect, and $\hat{\theta}_{t=1,\ldots, T}$ the sufficient statistics of the treatment effects reported on a cadence indexed by $t$ (e.g., daily estimates of the effects). We complete this specification with the likelihood $\mathcal{L}(\hat{\theta}_{t=1,\ldots, T}|\theta)$ and the prior $\pi(\theta)$, which jointly specify the posterior $\pi(\theta|\hat{\theta}_{t=1,\ldots, T})$. 

\subsection{Inference Based on Predictive Probabilities}
Our objective is to infer success before the experiment reaches its planned conclusion, i.e., at some analysis point $T' << T$. This is accomplished via the predictive probability of success (PPoS):

$$
PPoS = \int_{Y_{t=T', \ldots T}} \textbf{1}[\textbf{P}(\theta > 0|\hat{\theta}_{t=T',\ldots, T}) > (1-\alpha)]\pi(\hat{\theta}_{t=T',\ldots, T}|\hat{\theta}_{t=1,\ldots, T'})\mathrm{d}\hat{\theta}_{t=T',\ldots, T}.
$$
$\pi(\hat{\theta}_{t=T',\ldots, T}|\hat{\theta}_{t=1,\ldots, T'})$ characterizes the predictive distribution $\int_{\theta}\mathcal{L}(\hat{\theta}_{t=T',\ldots, T}|\theta)$ $ \pi(\theta|\hat{\theta}_{t=1, \ldots T'})\mathrm{d}\theta$.
One may stop the trial early if $PPoS>\gamma$ for some threshold $\gamma$. The following section demonstrates  how this quantity may be inferred via Monte-Carlo simulation.

\subsection{Estimation Strategy: Posterior Predictive Simulation}
We can estimate PPoS via simulation using algorithm~\ref{alg_ppos}, which draws from the work of~\citet{berry2010bayesian}. The objective is to utilize the predictive distribution at the interim point to generate `data' at the conclusion of the experiment, characterize the posterior at this hypothetical conclusion, and then average over the possible states.

\begin{algorithm}[h]
 Estimate $\hat{\theta}_{t=1, \cdots, T'}$:$\pi(\hat{\theta}_{t=T',\cdots, T}| \hat{\theta}_{t=1,\cdots, T'})$\\
     \For{$k \in 1,\cdots, K$}{
        Simulate $\hat{\theta}^{(k)}_{t=T',\cdots, T}$ from $\pi(\hat{\theta}_{t=T'\cdots, T}| \hat{\theta}_{t=1,\cdots, T'})$ \\
        Estimate $\pi(\theta^{(k)} | \hat{\theta}_{t=1,\cdots, T'}^{(k)})$. \\
    }
    Estimate $\widehat{PPoS} = \frac{1}{K} \sum_{k=1}^{K}[\textbf{P}(\theta^{(k)} > 0|\hat{\theta}_{t=1,\ldots, T'}) > (1-\alpha)] $.
 
\caption{Computing the PPoS via Simulation.}
\label{alg_ppos}
\end{algorithm}

\subsection{Model Specification}
As in~\citet{kessler2024overcoming}, we utilize a conjugate Gaussian parameterization of the Bayesian model which assumes a well behaved estimate of the scale parameter $\sigma^{2}$:
\begin{align*}
\hat{\theta}_{t=1,\ldots, T'}|\theta, \sigma^{2} &\sim \mathrm{N}(\theta, \sigma^{2}),\\
\theta|m_{0}, \tau &\sim \mathrm{N}(m_{0}, \tau).
\end{align*}

Under this specification, with no meaningful prior information at hand~\citep{berger2004case}, as $\tau \rightarrow \infty$, the posterior and predictive distributions become:
\begin{align*}
\theta | \hat{\theta}_{t=1\ldots, T'} &\sim \mathrm{N}\left(\frac{\sum_{t=1}^{T'}\hat{\theta}_{t}}{T'}, \frac{\sigma^{2}}{T'}\right), \\
\hat{\theta}_{t=T'\ldots, T}| \hat{\theta}_{t=1\ldots, T'} &\sim \mathrm{N}\left(\frac{\sum_{t=1}^{T'}\hat{\theta}_{t}}{T'}, \frac{\sigma^{2}}{T} + \frac{\sigma^{2}}{T'} \right).
\end{align*}

\subsection{Choice of Prior}
The choice of prior in this setting depends heavily upon the required properties of the system, like the desired Type I and Type II error rates, or more general operating characteristics~\citep{zhou2021bayesian}. For this paper, in order to mimic classical Frequentist properties, we use a Jeffreys prior~\citep{gelman1995bayesian}. This is particularly useful in the online experimentation domain, where practitioners fuse Bayesian analysis of experiments with Frequentist frameworks for decision making. In section~\ref{sec:discussion} we will discuss future research in this space including the development of \textit{informative} priors.

\subsection{Ongoing Validation via Predictive Checking}
Given the complexities of validation in experimental settings where there is no \textit{ground truth}, we take inspiration from \textit{predictive checking}~\citep{gelman1996posterior}.  
This type of assessment uses predictive simulation of new data $\hat{\theta}_{t=1, \ldots, T}^{rep}$ under a model of interest $M$, and a related statistic $f(\cdot)$, to characterize discrepancies against its observed counterpart $g(f(\hat{\theta}_{t=1, \cdots, T}^{rep}), f(\hat{\theta}_{t=1, \cdots, T}))$. This discrepancy can be used to assess any aspect of the model we want to validate (e.g., prior parameter choices or decision boundary), facilitated by the construction of a reference distribution:
\begin{align}
p[g(f(\hat{\theta}_{t=1, \cdots, T}^{rep}), f(\hat{\theta}_{t=1, \cdots, T}))|M, \hat{\theta}].
\end{align}
This flexible construction enables assessment of quantities like coverage for uncertainty intervals or differences between interim analyses and end of experiment outcomes -- the latter of which we do in the next section to gauge performance of the PPoS based decision rule.


\section{Results}
\label{sec:results}
\begin{figure}
    \centering
    \begin{subfigure}[t]{0.32\textwidth}
        \centering
        \includegraphics[width=\linewidth]{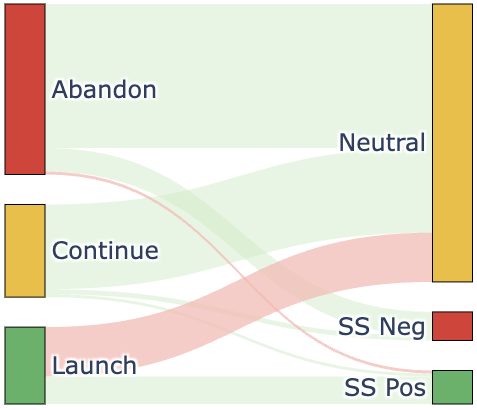} 
        \caption{Practical Heuristic} \label{fig:naive}
    \end{subfigure}
    \begin{subfigure}[t]{0.32\textwidth}
        \centering
        \includegraphics[width=\linewidth]{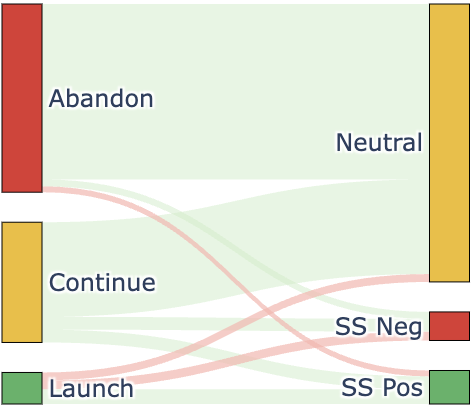} 
        \caption{Always Valid} \label{fig:av}
    \end{subfigure}
    \begin{subfigure}[t]{0.32\textwidth}
        \centering
        \includegraphics[width=\linewidth]{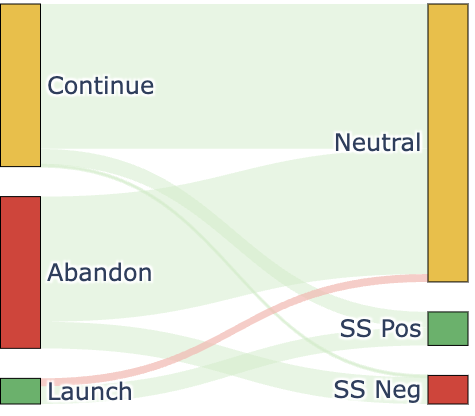} 
        \caption{PPoS} \label{fig:ppos}
    \end{subfigure}
    \caption{Comparison of decision rules.}
    \label{fig:comparison}
\end{figure}

We assess the performance of our proposed technique using 345 experiments conducted over a one month period. 
We compare three decision-making techniques:
\begin{itemize}
    \item \textbf{Practical Heuristic}: The experiment is abandoned as a failure if the lower endpoint of a 90\% uncertainty interval for the effect is below $l$, declared a success if it exceeds $m$, and continues the experiment otherwise.
    \item \textbf{Always Valid}: The experiment is abandoned as a failure if the always-valid p-value is greater than 0.95, declared a success if it is less than 0.05, and continues the experiment otherwise \citep{johari2022always}.
    \item \textbf{PPoS}: The experiment is abandoned as a failure if the PPoS is less than 0.1, declared a success if it is greater than 0.9, and continues the experiment otherwise.
\end{itemize}
Each interim analysis is conducted at day 7 of the experiment and compared against the end of experiment decision at day 14. Figure~\ref{fig:comparison} plots the results.

The practical heuristic correctly abandons 170 experiments as failures that would have been neutral or statistically significant negative if the experiment had been run to completion. However, 50 of the 78 experiments it suggests launching would have been neutral (i.e., false positives) if run to completion -- a common problem stemming from decisions using unadjusted peeking~\citep{berman2018p}.
Similarly, the Always-Valid rule correctly abandons 185 experiments that would have ultimately resulted in failures. It slightly improves the false positive rate relative to the naive rule: only 17 of the 32 experiments it suggests launching are neutral or statistically significant negative at completion. 

PPoS based decision correctly curtails 156 experiments It improves false positives even further: only 8 of the 26 experiments being recommended for launch are neutral at completion, with no negative findings. As it correctly stops a large number of experiments while maintaining the lowest false positive rate, PPoS based stopping is preferable in our case to both the practical heuristic, and the always-valid p-value.

\section{Discussion}
\label{sec:discussion}
We introduce Bayesian Predictive Probabilities for the interim analysis of Online Experiments as a statistically efficient and principled means of making decisions, along with a simulation based system for estimation and an ongoing validation strategy. We demonstrate, in accordance with the likelihood principle, that using a decision rule based on the predictive probability of success offers strong performance relative to both heuristic practices common in the online experimentation domain, and also to principled but inefficient Frequentist approaches. In specific, a PPoS based decision mechanism curtails a large collection of experiments while maintaining fidelity with end of experiment outcomes.

Systems based on predictive probabilities offer many avenues for further research. First, given the large slew of experiments from experienced analysts available at any given time, formal informative prior elicitation is possible~\citep{johnson2010methods}, and can improve expediency in decision making. One can leverage offline evaluation~\citep{Radwan2024Eval} and domain knowledge~\cite{LDP-BN} to construct such informative priors. Second, sign and magnitude errors~\citep{gelman2014beyond} remain a concern in low power settings -- an evergreen concern in the online experimentation domain. These are typically addressed by way of using \textit{shrinkage} priors.

\bibliographystyle{abbrvnat}
\bibliography{early_stopping}

\end{document}